\documentclass[iop,revtex4]{emulateapj}

\usepackage{color}
\usepackage{ulem}
\usepackage{url}

\shorttitle{Broad-band X-ray study of IC 342 X-1}
\shortauthors{Shidatsu et al.}

\begin{document}

\title{NuSTAR and Swift observations of the ultraluminous X-ray source IC 342 X-1 in 2016: witnessing spectral evolution} 

\author{M. Shidatsu\altaffilmark{1}, 
Y. Ueda\altaffilmark{2},
S. Fabrika\altaffilmark{3,4} 
}
\email{megumi.shidatsu@riken.jp}

\altaffiltext{1}{Institute of Physical and Chemical Research (RIKEN), 2-1 Hirosawa, Wako, Saitama 351-0198, Japan}
\altaffiltext{2}{Department of Astronomy, Kyoto University, Kitashirakawa-Oiwake-cho, Sakyo-ku, Kyoto 606-8502, Japan}
\altaffiltext{3}{Special Astrophysical Observatory, Nizhnij Arkhyz 369167, Russia}
\altaffiltext{4}{Kazan Federal University, Kremlevskaya 18, Kazan 420008, Russia}

\setcounter{footnote}{0}

\begin{abstract}
We report on an X-ray observing campaign of the ultraluminous 
X-ray source IC 342 X-1 with {\it NuSTAR} and {\it Swift} 
in 2016 October, in which we captured the very moment when  
the source showed spectral variation. The {\it Swift/XRT} 
spectrum obtained in October 9--11 has a power-law shape 
and is consistent with those observed in the coordinated 
{\it XMM-Newton} and {\it NuSTAR} observations in 2012.
In October 16--17, when the 3--10 keV flux became $\approx$4 
times higher, we performed simultaneous {\it NuSTAR} and 
{\it Swift} observations. In this epoch, the source showed a 
more round-shaped spectrum like that seen with {\it ASCA} 
23 years ago. Thanks to the wide energy coverage and high 
sensitivity of {\it NuSTAR}, we obtained hard X-ray data 
covering up to $\sim$30 keV for the first time during the high 
luminosity state of IC 342 X-1. The observed spectrum has 
a broader profile than the multi-color disk blackbody model. 
The X-ray flux decreased again in the last several hours of 
the {\it NuSTAR} observation, when the spectral shape 
approached those seen in 2012 and 2016 October 9--11.
The spectra obtained in our observations and in 2012 
can be commonly described with disk emission and its 
Comptonization in cool ($T_{\rm e} \approx 4$~keV), 
optically-thick ($\tau \approx 5$) plasma. The spectral 
turnover seen at around 5--10 keV shifts to higher 
energies as the X-ray luminosity decreases. This behavior 
is consistent with that predicted from recent numerical 
simulations of super-Eddington accretion flows with 
Compton-thick outflows. We suggest that the spectral 
evolution observed in IC 342 X-1 can be explained by 
a smooth change in mass accretion rate.
\end{abstract}


\section{Introduction}

\begin{deluxetable*}{lcccc}[htbp]
\tablecaption{Log of {\it Swift} and {\it NuSTAR} observations
\label{tab_obslog}}
\tablewidth{0pt}
\tablehead{
\colhead{ObsID} & \colhead{Start time (UT)} & \colhead{End time (UT)} & \multicolumn{2}{c}{Net exposure (ks)} 
}
\startdata
\multicolumn{1}{c}{\it NuSTAR} & & & FPMA & FPMB \\
%\tableline 
90201039002 & 2016 Oct. 16 02:11:08 & 2016 Oct. 17 02:51:08 & 49.1 & 49.0 \\
\tableline 
\multicolumn{1}{c}{{\it Swift}\footnotemark[a]} & & & \multicolumn{2}{c}{XRT} \\
%\tableline 
00031987018 & 2016 Oct. 09 03:42:38 & 2016 Oct. 09 05:34:52 
& \multicolumn{2}{c}{0.94} \\
00080321019 & 2016 Oct. 10 03:34:04 & 2016 Oct. 10 05:25:54 
& \multicolumn{2}{c}{0.94} \\
00080321020 & 2016 Oct. 11 03:31:38 & 2016 Oct. 11 05:20:55 
& \multicolumn{2}{c}{0.94} \\
00080321001 & 2016 Oct. 16 06:23:00 & 2016 Oct. 16 06:38:53 
& \multicolumn{2}{c}{0.94} \\
00080321002 & 2016 Oct. 16 10:54:26 & 2016 Oct. 16 11:10:54 
& \multicolumn{2}{c}{0.96} \\
00080321003 & 2016 Oct. 16 14:05:26 & 2016 Oct. 16 14:18:53 
& \multicolumn{2}{c}{0.78} \\
00080321004 & 2016 Oct. 16 18:51:30 & 2016 Oct. 16 19:08:53
& \multicolumn{2}{c}{1.0} \\
00080321005 & 2016 Oct. 16 22:03:30 & 2016 Oct. 16 22:20:54
& \multicolumn{2}{c}{1.0} \\
00080321006 & 2016 Oct. 17 01:18:33 & 2016 Oct. 17 01:34:53
& \multicolumn{2}{c}{0.96}
\enddata
\tablenotetext{a}{The XRT was operated in the Photon Counting 
(PC) mode in all the {\it Swift} observations.}
\end{deluxetable*}

Ultraluminous X-ray sources \citep[ULX: ][]{mak00}, 
found in off-nuclear regions of nearby galaxies, 
have luminosities exceeding the Eddington 
limit for stellar-mass black 
holes: $\sim 10^{39}$ erg s$^{-1}$ \citep{fab89}. 
The mass of the compact object and the mass accretion 
rate of ULXs have been controversial. There have been 
two major ideas present: (1) an intermediate mass 
($\gtrsim 100 M_\sun$) black hole accreting at 
sub-Eddington rates \citep[e.g.,][]{col99,mak00,mil03} 
and (2) a smaller mass black hole: 
a stellar-mass (up to a few ten $M_\sun$) black hole 
accreting at super-Eddington rates 
\citep[e.g.,][]{fab01,kin01,wat01,ebi03,pou07}
or a black hole with several ten $M_\sun$ 
produced from low metallicity stars 
\citep[e.g.,][]{map09,bel10} accreting at or above 
Eddington rates.
If the former is the case, they could be the long-sought 
missing link between stellar-mass black holes and 
supermassive black holes in the center of galaxies. 
If the latter is the case, they would be the best 
laboratory to study super-Eddington accretion. In 
any case, ULXs are important objects in many aspects, 
including physics of black hole accretion, and 
formation/growth of binary systems and black holes 
themselves. 

While extremely luminous ULXs are suggested to harbor 
an intermediate mass black hole accreting at sub-Eddington 
rates \citep{far09,sut12}, the dominant population, with 
luminosities just above $10^{39}$ erg s$^{-1}$ up to 
$\sim 10^{40}$ erg s$^{-1}$ \citep{swa11, wal11}, 
encompasses the most 
promising candidates of stellar-mass, super-Eddington 
accretors. Signs of massive outflows, predicted by 
numerical simulations of super-Eddington accretion 
flows \citep{ohs05, ohs11}, have been found directly 
and indirectly. \citet{mid14} suggested that the residuals 
seen in many ULX spectra can be explained by absorption 
structures of ultrafast disk winds with a velocity of 
$\sim 0.1c$. Indeed, absorption features of such 
massive winds have now been discovered at $\sim$1 keV in the 
ULX NGC 1313 X-1 \citep{pin15}, although the statistical 
significance is still not very high. Moreover, \citet{wal16} 
have recently detected a weak Fe K$\alpha$ absorption 
line at 8.8 keV in the same object and estimated the outflow 
velocity of $\sim 0.2$c, which is consistent with that 
obtained in \citet{pin15}. 
Deep optical spectroscopy of several ULXs also suggests 
the existence of disk winds \citep{fab15}. These results 
support that super-Eddington accretors are actually 
present in the ULXs. In addition, X-ray pulsars have 
recently been detected in a few ULXs 
\citep{bac14, isr16, fue16, isr17}, and at least 
one of them shows a typical X-ray spectrum of 
the known ULXs.
This may suggest that, although only a small 
number of the detections have been present so far, 
neutron star binaries at super-Eddington 
luminosities constitute a substantial 
population of ULXs.

To understand the physical origin of the strong 
radiation from ULXs, Galactic black hole X-ray 
binaries (BHXBs) can be used for comparison 
as a template of accreting stellar-mass black 
holes below Eddington luminosity. 
Thanks to their proximity, BHXBs are very bright 
in their outbursts and have provided high-quality 
broad-band X-ray data. They show several distinct 
spectra at different luminosities 
\citep[see][for reviews]{mcc06,don07}. 
The two states that are most often seen are the 
low/hard state, with a hard power-law spectra 
with an exponential cutoff at $\sim 100$ keV, 
and the high/soft state, in which a soft X-ray 
emission from the standard accretion disks 
\citep{sha73} dominates the spectrum.  

Spectral variability of ULXs has been studied 
by compiling data from snapshot observations 
\citep[e.g.,][]{pin12,pin14,wal14}. The results 
show that ULXs exhibit a variety of X-ray spectra 
\citep{sut13} but have some distinct characteristics 
in their spectral profiles compared with those seen 
in Galactic BHXBs. 
They usually have a spectral turnover at 
energies from a few keV to just above 10 keV 
\citep[e.g.,][]{gla09}, much lower than that 
in the low/hard state in Galactic BHXBs. 
For some ULXs, the inner disk temperature 
($T_{\rm in}$) estimated from their spectra 
anti-correlates with the luminosity ($L$), 
instead of the $L \propto T_{\rm in}^4$ relation, 
which usually holds in the high/soft state 
of BHXBs (\citealt{fen09}; \citealt{kaj09}; 
but see \citealt{mil13}). 
These differences may suggest that ULXs have 
different properties of accretion flows 
from those of Galactic BHXBs, although more 
X-ray observations, in particular during state 
transitions, would be needed for a deeper 
understanding of the origin of such variability.

In 2016 October, we carried out an X-ray observing 
campaign of the ULX IC 342~X-1 with 
{\it Nuclear Spectroscopic Telescope Array} 
\citep[{\it NuSTAR};][]{har13} and {\it Swift} \citep{geh04}, 
in which we successfully monitored its 
spectral evolution. IC 342 X-1 has been observed 
quite intensively in the 
X-ray band. Comparing {\it ASCA} data taken in 1993 
and 2000, \citet{kub01} reported a transition 
between ``the disk dominant state'' at a high 
luminosity and ``the power-law spectral state'' 
at a low luminosity. Spectral variation was 
observed in the former state, which was interpreted 
as a change in the inner disk radius and 
temperature \citep{miz01}. 
A similar round-shaped spectrum was detected 
with {\it Chandra} as well \citep{mar14}. 
\citet{yos13} found two distinct power-law 
spectral states at different luminosities, 
which showed a spectral turnover at 
different energies. The lower-luminosity 
power-law state was studied more deeply with 
high-quality broad-band X-ray data covering 
up to $\approx30$ keV with coordinated 
XMM-Newton and NuSTAR observations in 2012 
\citep{ran15}, which confirmed the presence of 
the spectral turnover at $\approx 10$ keV.

In this paper, we report the results of the observations 
in 2016 October. Section~\ref{sec_obsdat} describes 
the details of the observations and the data reduction. 
Section~\ref{sec_spec} presents spectral analysis that 
we performed. In Section~\ref{sec_discussion} we discuss 
possible physical interpretations of the results and the 
mass of the compact object in IC 342 X-1.
In all sections, errors represent the 90\% confidence 
range for a single parameter, unless otherwise stated. 
We used HEASOFT version 6.19 for data reduction, and XSPEC 
version 12.9.0n for spectral analysis. 
Throughout the article, we refer to the table given by 
\citet{wil00} as the solar abundances. We assumed 
the distance to the galaxy IC 342 as 3.93 Mpc \citep{tik10}.

\section{Observations and Data Reduction} \label{sec_obsdat}

We performed {\it NuSTAR} time-of-opportunity (ToO) 
observation of IC 342 X-1 in 2016 October 16--17 
with a net exposure of $\approx$50 ksec. We also 
carried out {\it Swift} ToO observations 9 times 
in 2016 October, each of which have a net exposure 
of $\sim$1 ks. Six of them were performed in $\approx$2-hour 
intervals during the {\it NuSTAR} observation and 
the rest were on October 9, 10, and 11. A log of 
the observations is given in Table~\ref{tab_obslog}.

The {\it NuSTAR} data were processed by using the 
tools {\tt nupipeline} included in the NuSTAR Data 
Analysis Software ({\tt nustardas}) version 1.6.0 
with the calibration database (CALDB) version 20160922. 
The source and background extraction regions were defined 
as circular regions with a radius of 50'' centered on the 
target position (which is the same as \citealt{ran15}) 
and with a radius of 80'' in a nearby source free 
region on the same detector, respectively. 
{\it NuSTAR} light curves for each of 
focal plane modules (FPMA and FPMB) were produced through 
{\tt nuproducts} with standard settings for point 
sources described in the NuSTAR analysis 
guide\footnote{http://heasarc.gsfc.nasa.gov/docs/nustar/analysis/nustar\_sw\\guide.pdf}. 
The {\it Swift}/XRT data were reprocessed with 
{\it Swift} CALDB version 20160706 through the 
pipeline tool {\tt xrtpipeline}. The source events were extracted 
from a circular region centered on the source position with 
a radius of 30'' and background events were from a circle 
with a radius of 4.8' in a nearby source-free field 
on the same detector.

\begin{figure}
\plotone{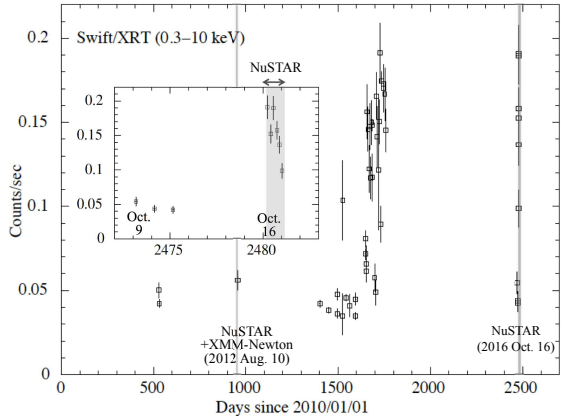}
\caption{
{\it Swift}/XRT 3--10 keV light curve of IC 342 X-1 with 
one point per observation. The {\it NuSTAR} observation 
performed in 2016 October 16--17 and the coordinated 
{\it XMM-Newton}$+${\it NuSTAR} observation on 2012 
August 10 are also indicated as grey shaded regions. 
The inset is an enlarged view focusing on the period of 
our observing campaign in 2016. 
\label{fig_swiftLC}
}
\end{figure}

\begin{figure}
\plotone{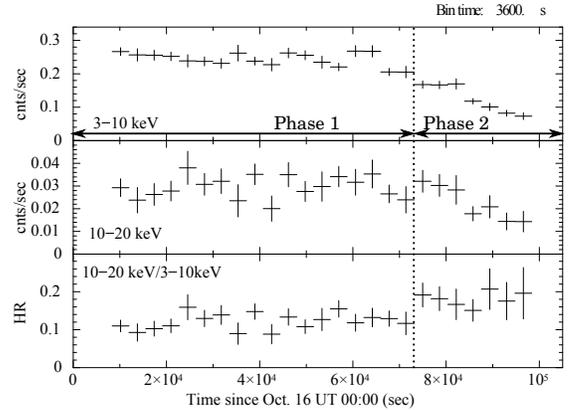}
\caption{
{\it NuSTAR}/FPMA light curves of IC 342 X-1 in the 
3--10 keV (top) and 10--20 keV bands (middle) and 
the hardness ratios (bottom). The observation period 
is divided into two phases: MJD 57677.094--57677.847 
(Phase 1) and MJD 57677.847--57678.347 (Phase 2), 
which are used to produce X-ray spectra separately  
for different hardness ratios. 
\label{fig_nustarLC}
}
\end{figure}

Figure~\ref{fig_swiftLC} shows the {\it Swift}/XRT 
light curve of IC 342 X-1 in the 0.3--10 keV band for 
the past $\approx$6 years. This was made with 
the online data analysis tools\footnote{http://www.swift.ac.uk/user\_objects/} 
provided by the UK 
{\it Swift} Science Data Centre at the University 
of Leicester. The count rate varied by a 
factor of $\approx$4. The source was in the lowest 
flux state on 2016 October 9, 10, and 11, whereas 
the flux reached the maximum level in our {\it NuSTAR} 
observation performed about 1 week after.
Figure~\ref{fig_nustarLC} presents the 
background-subtracted {\it NuSTAR} light curves and 
the hardness ratio. The source kept a fairly constant 
X-ray flux during the first two thirds of the entire 
period in the {\it NuSTAR} observation, and afterwards 
it gradually become fainter and harder.

To study the spectral evolution, we split  
the {\it NuSTAR} observation into two phases 
based on the count rate and hardness ratio 
(see Fig~\ref{fig_nustarLC}): 
MJD 57677.094--57677.847 
(hereafter Phase 1) 
and MJD 57677.847--57678.347 (Phase 2). 
We created the time-averaged 
{\it NuSTAR} and {\it Swift}/XRT spectra for each 
of them. 

{\it NuSTAR} spectra and response files for the 
individual phases were produced with {\tt nuproducts}.
We combined all the {\it Swift}/XRT data in the same 
phase. Four of six datasets taken in October 16
(ObsID 00080321001, 00080321002, 00080321003, and 
00080321004) and two of them taken in October 
16--17 (ObsID 00080321005 and 00080321006) were 
co-added to produce the time-averaged 
XRT spectra for Phase 1 and Phase 2, respectively. 
The redistribution matrix file (RMF) 
{\tt swxpc0to12s6\_20130101v014.rmf} was used for the   
XRT spectra. The auxiliary response file (ARF) for each 
phase was generated via {\tt xrtmkarf} by using a 
combined exposure map created from the data of all 
ObsIDs for the same phase. The time-averaged XRT 
spectrum for October 9--11 and its ARF were also made 
in the same manner by combining the three observations 
in that period. 

For comparison, we also reduced the {\it ASCA} 
data taken in 1993, and the simulnatenous 
{\it XMM-Newton} and {\it NuSTAR} data 
in 2012 August 10 (ObsID$=$0693850601 and 30002032003, 
respectively; see \citealt{ran15} for more details 
of the observations) and created the time-averaged 
broad-band X-ray spectra. 
The {\it ASCA} data were reduced in the same 
manner as \citet{oka98} by using the archival data 
and the {\it ASCA} CALDB downloaded from the HEASARC 
online CALDB page\protect\footnotemark[6] on 2016 December 10.
For the {\it XMM-Newton} data, we utilized the Science 
Analysis System (SAS) version 15.0.0 and the latest Current 
Calibration File (CCF) as of 2016 October 23 to reprocess  
them and produce the spectrum. We excluded high 
background intervals following the {\it XMM-Newton} ABC 
Guide\footnote{http://heasarc.gsfc.nasa.gov/docs/xmm/abc/} 
and selected events with 
\verb+FLAG==0 && PATTERN <= 4+ for EPIC-pn and
with \verb+FLAG==0 && PATTERN <= 12+ for the 
two EPIC-MOS cameras.
The source and background events were extracted 
from a circular region with a radius of 30'' 
centered on the source position and with 
a radius of 60'' in a nearby source-free region 
on the same detector, respectively. The RMF and ARF 
were generated with the SAS tools {\tt rmfgen}, and 
{\tt arfgen}, respectively.
The {\it NuSTAR} data were reduced in the same 
manner with the same software as the 2016 data. 

The {\it NuSTAR} FPMA and FPMB spectra were combined 
by using {\tt addascaspec} to improve 
statistics and used throughout the following spectral 
modelling. We have confirmed that the results remain 
unchanged within the 90\% confidence ranges 
if the two spectra are analyzed separately. 
Considering the low statistics and relatively large 
uncertainty of the {\it Swift}/XRT data, we fixed the 
cross normalizations between the {\it Swift}/XRT and 
{\it NuSTAR} data for the two phases at $1:1$. We 
confirmed that their normalizations for Phase 1 
and for Phase 2 were consistent {within uncertainties 
at $\lesssim 10$\% and $\lesssim 20$\% levels}, respectively. 
For 2012 data, the normalizations of the 
{\it XMM-Newton}/EPIC-pn, EPIC-MOS1, and 
EPIC-MOS2 spectra were found to be consistent 
with each other at $\sim 3$\% levels, while a 
$\sim 10$\% difference was detected between those 
of the {\it XMM-Newton} and {\it NuSTAR} spectra. 
Thus we set the cross normalization among the 
cameras of {\it XMM-Newton} as $1:1:1$ and varied 
that between the {\it XMM-Newton} and {\it NuSTAR}.

\begin{figure}
\plotone{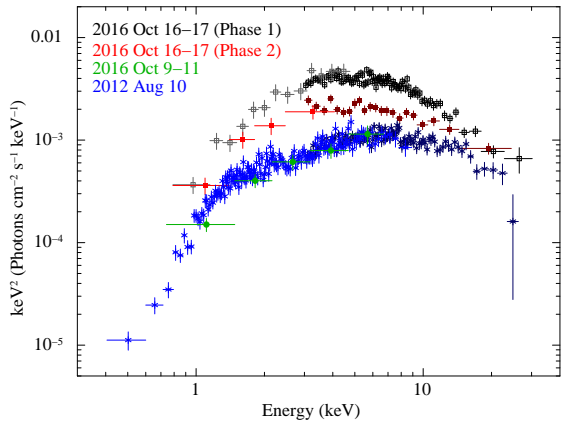}
\caption{
Time-averaged spectra of IC 342 X-1 in different epochs. 
Black open squares and red filled squares present the 
simultaneous {\it Swift}/XRT and {\it NuSTAR} spectra in the two 
phases defined in Figure~\ref{fig_nustarLC}. The {\it Swift}/XRT 
spectrum in 2016 October 9--11 is plotted in green (filled circle), 
and the {\it XMM-Newton}$+${\it NuSTAR} data taken in 2012 August 10 
are shown in blue (cross). The darker colors indicate the {\it NuSTAR} 
data and the brighter colors correspond to the {\it Swift} or {\it XMM-Newton} 
data. Only EPIC-MOS1 data are shown for the {\it XMM-Newton} spectrum. 
FPMA and FPMB data are combined in all NuSTAR spectra. 
\label{fig_compare_spec}}
\end{figure}

\begin{figure}
\plotone{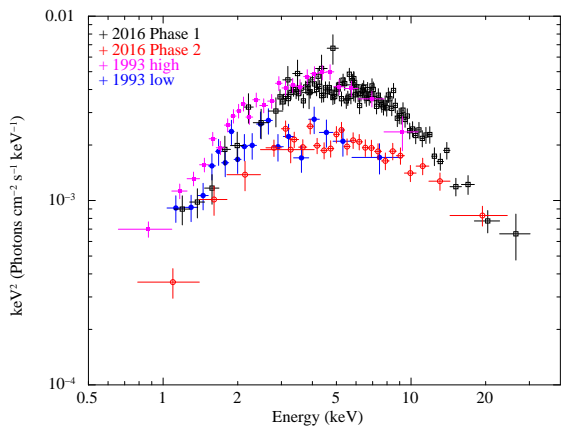}
\caption{
Comparison of the spectra in Phase 1 (black open squares) and 
Phase 2 (red open circles), and the {\it ASCA}/GIS2 spectra 
obtained in 1993 during the high-flux periods (pink filled squares) 
and the low-flux periods (blue filled circles) defined in 
\citet{miz01}. 
\label{fig_compare_spec_asca}}
\end{figure}

\footnotetext[8]{http://heasarc.gsfc.nasa.gov/docs/heasarc/caldb/caldb\_\\supported\_missions.html}
\setcounter{footnote}{8}

\section{Spectral Analysis and Results} \label{sec_spec}

\subsection{Long-term Spectral Variation} 
In Figure~\ref{fig_compare_spec}, we present the 
time-averaged {\it NuSTAR} and {\it Swift}/XRT 
spectra on different occasions in 2016 October. 
For comparison, we also plot the simultaneous 
{\it NuSTAR}$+${\it XMM-Newton} spectra obtained 
on 2012 August 10.
The spectrum in 2016 October 9--11 is fully 
consistent with that obtained in 2012, suggesting that 
the source was in the same state during the two epochs. 

About 1 week after the {\it Swift/XRT} observations,
the source increased its luminosity and changed 
the profile of the continuum spectrum.
In Phase 1, it marked the highest 3--20 keV flux in 
Figure~\ref{fig_compare_spec}. It showed a more 
round shaped 
spectrum, which start bending at a somewhat lower 
energy than the spectrum in 2012. 
The Phase-1 spectrum below $\approx$10 keV looks 
like that in the ``soft state'' observed with 
{\it ASCA} in 1993 \citep{kub01,miz01}, which 
was well reproduced with a single multi-color 
disk blackbody model \citep[MCD model;][]{mit84}. 
In Figure~\ref{fig_compare_spec_asca}, we compare 
the {\it ASCA} spectra and the {\it NuSTAR}$+${\it Swift} 
spectra. 
We find that the Phase-1 spectrum has a 
similar continuum profile to that obtained during 
the high-flux periods \citep{miz01} in the 
{\it ASCA} observation.

As the X-ray flux decreased from Phase 1 to Phase 2, 
the spectral shape became consistent with that in 
the low-flux periods of the {\it ASCA} observation 
in 1993 (see Fig.~\ref{fig_compare_spec_asca}) and 
got closer to what was seen in 2012. 
The Phase-2 spectrum has a spectral turnover at 
a similar energy to the Phase-1 spectrum, but 
below that energy, 
it appears to have a somewhat flatter profile 
in the $\nu F\nu$ form, like the 2012 data 
(Fig.~\ref{fig_compare_spec}). 
In the following sections, we attempt to assess 
the differences/similarities in continuum profile 
among the two phases and the 2012 epoch more 
quantitatively.

\begin{figure}
\plotone{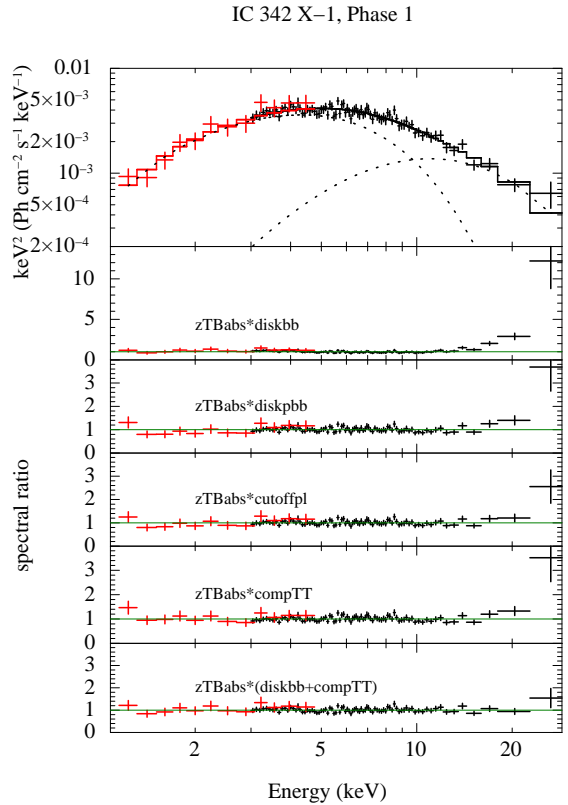}
\caption{Results of spectral fit for Phase 1 with various models.
{\it Swift} and {\it NuSTAR} data are shown in red and black, 
respectively. The top panel presents the Phase-1 spectrum and 
the best-fit {\tt zTBabs*(diskbb+compTT)} model. The dotted 
lines show the contributions of the {\tt diskbb} component 
(with a lower peak energy) and the {\tt compTT} component 
(higher peak energy).
The rest panels plot the data-to-model ratios for the {\tt zTBabs*diskbb}, 
{\tt zTBabs*diskpbb}, {\tt zTBabs*cutoffpl}, {\tt zTBabs*compTT}, 
and {\tt zTBabs*(diskbb+compTT)} models, from top to bottom.
\label{fig_specfit}}
\end{figure}

\begin{deluxetable*}{lcccc}
\tablecaption{Best-fit parameters of various spectral 
models for the {\it NuSTAR}$+${\it Swift} data in Phase 1 
and Phase 2 and the {\it NuSTAR}$+${\it XMM-Newton} data 
in 2012 \label{tab_specfit}}
\tablewidth{0pt}
\tablehead{
\colhead{Parameter} & \colhead{Unit} & \colhead{Phase 1} 
& \colhead{Phase2} & \colhead{2012 August 10} 
}
\startdata
\multicolumn{5}{l}{Model: zTBabs*diskbb} \\
\tableline 
$N_{\rm H}$ & $10^{22}$ cm$^{-2}$ & $0.02^{+0.09}_{-0.02~{\rm (pegged)}}$
& 0.0 (fixed) & $2^{+9}_{-2~{\rm (pegged)}} \times 10^{-3}$ \\
$kT_{\rm in}$ & keV & $2.35 \pm 0.05$ & $2.51^{+0.13}_{-0.12}$ 
& $2.40 \pm 0.05$ \\
$N_{\rm diskbb}$ & & $(2.0 \pm 0.2) \times 10^{-2}$  
& $(8.4 \pm 0.2) \times 10^{-3}$ & $5.1^{+0.5}_{-0.4} \times 10^{-3}$ \\
$\chi^2/$d.o.f. & & $176/94 = 1.87$ & $70/24 = 2.91$ & $1965/716 = 2.74$ \\
Flux & $10^{-11}$ erg cm$^{-2}$ s$^{-1}$ & $1.3$ & $0.71$ & $0.37$ \\
Luminosity & $10^{40}$ erg s$^{-2}$& $2.4$ & $1.3$ & $0.68$ \\
\tableline 
\multicolumn{5}{l}{Model: zTBabs*diskpbb} \\
\tableline 
$N_{\rm H}$ & $10^{22}$ cm$^{-2}$ & $0.8^{+0.3}_{-0.2}$ & $0.8$ (fixed) 
& $0.27 \pm 0.02$ \\
$kT_{\rm in}$ & keV  & $3.1 \pm 0.2$ & $4.0^{+0.4}_{-0.3}$ 
& $4.6^{+0.3}_{-0.2}$ \\
$p$ &  & $0.52^{+0.02}_{-0.02~{\rm (pegged)}}$ 
& $0.50^{+0.01}_{-0.00~{\rm (pegged)}}$ & $0.544 \pm 0.005$ \\
$N_{\rm diskbb}$ & & $2.1^{+1.2}_{-0.7} \times 10^{-3}$ 
& $3.5^{+1.9}_{-1.2} \times 10^{-4}$ 
& $1.4^{+0.4}_{-0.3} \times 10^{-4}$\\
$\chi^2/$d.o.f. & & $100/93 = 1.08$ & $20/23 = 0.86$ & $846/715 = 1.18$ \\
Flux & $10^{-11}$ erg cm$^{-2}$ s$^{-1}$ & $2.4$ & $1.4$ & $0.55$ \\
Luminosity & $10^{40}$ erg s$^{-2}$& $4.3$ & $2.6$ & $1.0$ \\
\tableline 
\multicolumn{5}{l}{Model: zTBabs*cutoffpl} \\
\tableline 
$N_{\rm H}$ & $10^{22}$ cm$^{-2}$ & $0.6 \pm 0.2$ & 0.6 (fixed) 
& $0.24 \pm 0.02$ \\
$\Gamma$ &  & $1.15 \pm 0.20$ & $1.65^{+0.19}_{-0.21}$ 
& $1.46 \pm 0.05$ \\
$E_{\rm cut}$ & keV & $4.9^{+0.7}_{-0.6}$ & $9.0^{+3.3}_{-2.1}$ 
& $11.8^{+1.4}_{-1.2}$ \\
$N_{\rm cutoffpl}$ & photon keV$^{-1}$ cm$^{-2}$ s$^{-1}$ 
& $2.9^{+0.6}_{-0.5} \times 10^{-3}$ & $(2.1 \pm 0.4) \times 10^{-3}$
& $6.7^{+0.4}_{-0.3} \times 10^{-4}$ \\
$\chi^2/$d.o.f. & & $88/93 = 0.95$ & $17/23 = 0.74$ 
& $883/715 = 1.24$ \\
Flux & $10^{-11}$ erg cm$^{-2}$ s$^{-1}$ & $1.8$ & $1.2$ 
& $0.55$ \\
Luminosity & $10^{40}$ erg s$^{-2}$& $3.4$ & $2.3$ & $1.0$ \\
\tableline 
\multicolumn{5}{l}{Model: zTBabs*compTT} \\
\tableline 
$N_{\rm H}$ & $10^{22}$ cm$^{-2}$ & $0.0^{+0.1}_{-0.0~{\rm (pegged)}}$ 
& 0 (fixed) & $0.22 \pm 0.03$ \\
$kT_{\rm 0}$ & keV & $0.63 \pm 0.06$ & $0.54^{+0.08}_{-0.07}$ 
& $0.20 \pm 0.02$ \\
$kT_{\rm e}$ & keV & $2.8^{+0.3}_{-0.2}$ & $3.8^{+1.4}_{-0.7}$ 
& $3.2^{+0.2}_{-0.1}$ \\
$\tau$ & & $5.4 \pm 0.6$ & $4.4^{+1.1}_{-0.9}$ 
& $6.4 \pm 0.3$ \\
$N_{\rm compTT}$ & & $1.1^{+0.1}_{-0.2} \times 10^{-3}$ 
& $(4.8 \pm 0.1) \times 10^{-4}$ & $4.7^{+0.5}_{-0.4} \times 10^{-4}$ \\
$\chi^2/$d.o.f. & & $97/92 = 1.05$ & $17/22 = 0.77$ & $758/714 = 1.06$ \\
Flux & $10^{-11}$ erg cm$^{-2}$ s$^{-1}$ & $1.3$ & $0.81$ & $0.48$ \\
Luminosity & $10^{40}$ erg s$^{-2}$& $2.4$ & $1.5$ & $0.89$ \\
\tableline 
\multicolumn{5}{l}{Model: zTBabs*(diskbb+compTT)}\tablenotemark{a} \\
\tableline  
$N_{\rm H}$ & $10^{22}$ cm$^{-2}$ & $0.2 \pm 0.1$ 
& 0.2 (fixed) & $0.56 \pm 0.08$ \\
$kT_{\rm in}$ & keV & $1.75^{+0.09}_{-0.02}$ & $0.6^{+1.2}_{-0.5}$ 
& $0.20 \pm 0.02$ \\
$N_{\rm bb}$ &  & $6^{+9}_{-2} \times 10^{-2}$  
& $0.5^{+0.5}_{-0.5~{\rm (pegged)}}$
& $1.5^{+1.7}_{-0.8} \times 10^2$ \\
$kT_{\rm e}$ & keV  & $> 3.0~(3.9)$\tablenotemark{b}
& $3.7^{+1.4}_{-1.1}$ & $3.2 \pm 0.2$ \\
$\tau$ &  & $>0.08~(5.8)$\tablenotemark{b} 
& $>3.4~(4.5)$\tablenotemark{b} & $6.3^{+0.4}_{-0.3}$ \\
$N_{\rm compTT}$ &  & $(1.0^{+9.9}_{-0.8}) \times 10^{-4}$ 
& $(4 \pm 3) \times 10^{-4}$ 
& $(4.9 \pm 0.5) \times 10^{-4}$ \\
$\chi^2/$d.o.f. & & $81/91 = 0.89$ & $17/21 = 0.81$ 
& $709/713 = 0.995$ \\
Flux & $10^{-11}$ erg cm$^{-2}$ s$^{-1}$ & $1.5$ & $0.89$ 
& $0.82$ \\
Luminosity & $10^{40}$ erg s$^{-2}$& $2.7$ & $1.6$ & $1.5$
\enddata
\tablecomments{All the fluxes and luminosities listed above are unabsorbed 
values estimated in the 0.3--30 keV band. In all models, we included 
an additional {\tt TBabs} component with $N_{\rm H} = 3.0 \times 10^{21}$ cm$^{-2}$ 
as the Galactic absorption, which is omitted in the table.}
\tablenotetext{a}{The seed temperature $kT_0$ of {\tt compTT} is 
linked to the inner disk temperature $kT_{\rm in}$ of {\tt diskbb}.}
\tablenotetext{b}{The numbers in parentheses are the best 
values that minimize $\chi^2$.}
\end{deluxetable*}

\subsection{Modelling the Phase-1 Spectrum} \label{sec_result_phase1}
To begin with, we focus on the Phase-1 data, 
which have much better statistics than the 
Phase-2 data. We perform spectral fitting using several 
models frequently adopted in previous studies 
of ULXs, and thereby describe the continuum 
profile in the high luminosity state of IC 342 X-1. 
The resultant parameters for each model are summarized 
in Table~\ref{tab_specfit} and the residuals of 
the fit are presented in Figure~\ref{fig_specfit}. 
In all models, we consider the Galactic absorption 
with an equivalent hydrogen column density of 
$N_{\rm H} = 3.0 \times 10^{21}$ cm$^{-2}$ (which is 
estimated from the H$_{\rm I}$ all-sky map by 
\citet{kal05} via the ftool command {\tt nh}), in 
addition to the neutral absorption in the binary 
system and the host galaxy with a redshift of 
$z=0.00095$. We use {\tt TBabs} and {\tt zTBabs} 
for the Galactic and local absorptions, 
respectively, and vary $N_{\rm H}$ of the latter 
as a free parameter. 

We first investigate if a single MCD model can 
reproduce the observed spectrum as in the case 
of the {\it ASCA} data \citep{kub01}. We find 
that this model does not produce an acceptable 
fit ($\chi^2/{\rm d.o.f.} = 176/94$). 
As shown in Figure~\ref{fig_specfit}, there 
remains significant deviation between the data 
and the model at the highest energies, 
suggesting that the observed continuum profile 
is broader than the MCD spectrum. 

A wider disk spectrum is realized in the case of 
the so-called ``slim disk'', which is believed to 
form at super-Eddington accretion rates \citep{abr88}. 
A slim disk has a smaller index $p$ of the radial 
dependence of the temperature than that for the 
standard accretion disks and thus produces 
a more broadened spectral profile. 
We replace the MCD component with 
{\tt diskpbb} \citep{min94}, where $p$ can vary, 
and fit the Phase-1 spectrum. The quality of fit 
is significantly improved from the result of the 
MCD model and become an acceptable level 
($\chi^2/{\rm d.o.f.} = 100/93$). Although the 
structures of residuals above 10 keV in 
Figure~\ref{fig_specfit} are quite evident, 
it is much less so for those below $\approx 2$ keV.

The spectral turnover seen at $\sim 10$ keV in ULX 
spectra are often explained by an optically-thick 
thermal Comptonization in the corona above the 
accretion disks or in strong outflows suggested 
to be driven from supercritical accretion flows 
\citep[e.g.,][]{ohs05}. 
We next investigate the possibility that such a 
Comptonized spectrum can reproduce the data. 
We apply two different models individually: 
a cutoff power-law ({\tt cutoffpl}) model as an 
approximation of a Comptonized 
spectrum, and the {\tt compTT} model \citep{tit94} 
as a more physical model. 
The latter model produces thermal Comptonized 
spectrum employing the Wien profile as the energy 
distribution of the seed photons. We leave the 
optical depth and the seed photon temperature 
as free parameters. Both models yield acceptable 
fits, but the structure of residuals similar to 
the {\tt diskpbb} model remain both in the 
{\tt cutoffpl} and {\tt compTT} models 
(see Fig.~\ref{fig_specfit}).

The single {\tt compTT} model assumes that the inner disk 
region emitting X-rays is fully obscured by the surrounding  
corona or outflows. To account for the possibility that the 
direct emission from the disk is visible outside the Comptonizing 
gas, we add a {\tt diskbb} component and link its inner disk 
temperature ($T_{\rm in}$) and the seed-photon temperature 
($T_{\rm 0}$) of the {\tt compTT} component. We find that this 
model can better reproduce the overall profile of the Phase-1 
spectrum than the single {\tt compTT} model. The $\chi^2$
value is reduced, 
and the discrepancies between the data and model are mitigated.

\subsection{Modelling the spectra in Phase 2 and in 2012} 
\label{subsec_phase2_2012_fit}

Using the models employed above, we next 
analyze the Phase 2 spectrum to characterize 
its continuum shape. 
For comparison, we also apply the same models 
to the 2012 spectrum, which is one of the 
highest quality data with the widest energy 
coverage among the low-luminosity state spectra 
of IC 342 X-1. The resultant parameters for each 
model in each epoch are listed in 
Table~\ref{tab_specfit}. For Phase 2, the parameters 
of {\tt diskbb}, the seed temperature of {\tt comptt}, 
and $N_{\rm H}$ are found to degenerate strongly 
due to poor statistics of the XRT data below 
$\approx 3$~keV. We thus fix $N_{\rm H}$ at the 
best-fit value for Phase 1 obtained with the 
same model. 

We find that the Phase 2 and 2012 spectra, 
as well as that of Phase 1, cannot be 
reproduced with the single {\tt diskbb} model. 
Comparing the resultant parameters of 
the phenomenological {\tt cutoffpl} model, 
a remarkable difference can be seen between 
Phase 1 and Phase 2: the photon index 
($\Gamma$) and the cutoff energy ($E_{\rm cut}$) 
are significantly larger in the latter than 
the former. We note that these differences 
are significant even if we vary $N_{\rm H}$ 
in Phase 2. The values of $\Gamma$ and 
$E_{\rm cut}$ in Phase 2 are consistent with 
those estimated for the 2012 spectrum. 
This suggests that the spectral profile 
became closer to that of the 2012 spectrum 
as the X-ray flux decreased  
from Phase 1 to Phase 2. In the single 
{\tt compTT} and {\tt diskbb+compTT} models, 
however, we do not detect any significant 
differences in the values of $T_{\rm e}$ 
and $\tau$.

\begin{figure}
\plotone{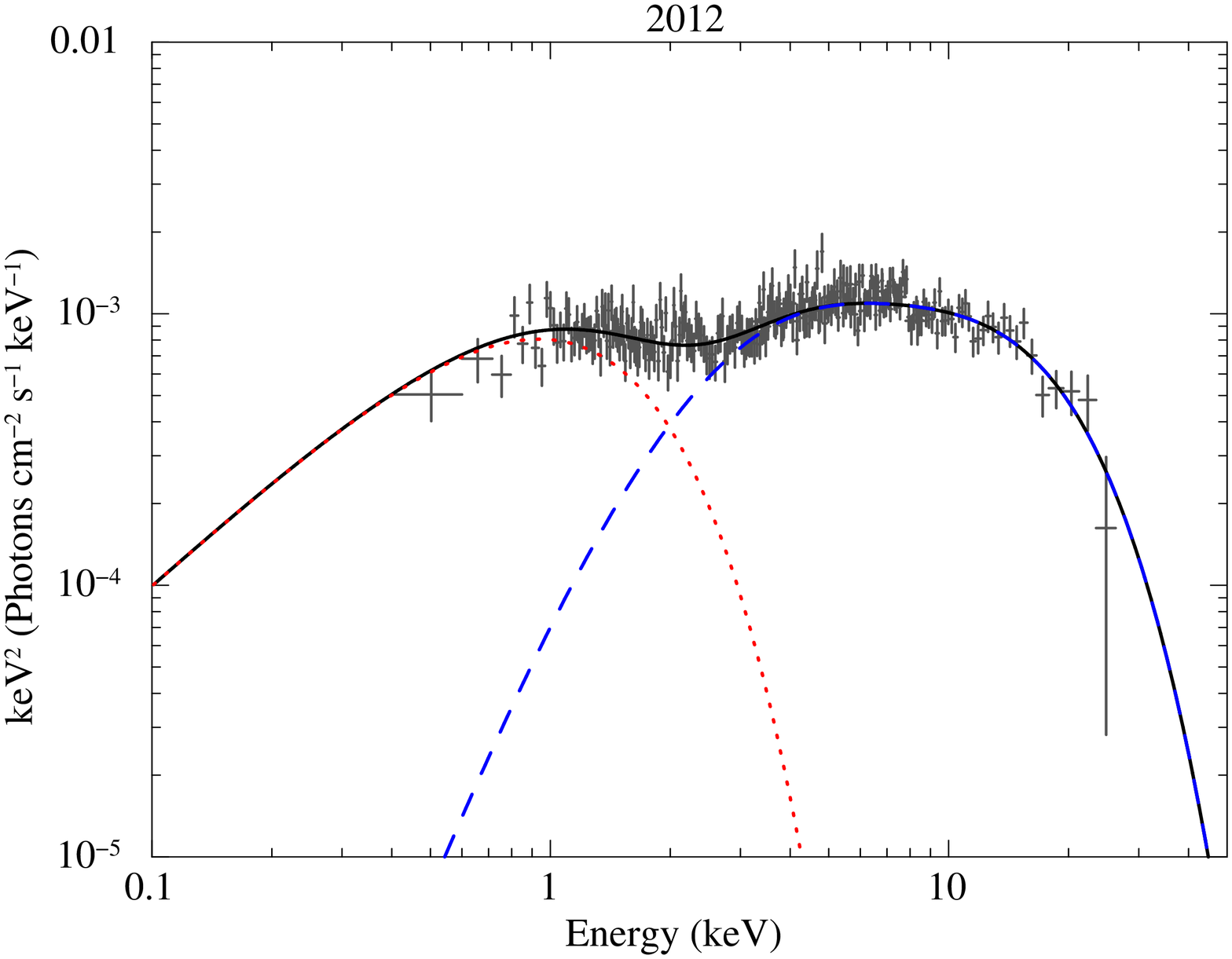}
\plotone{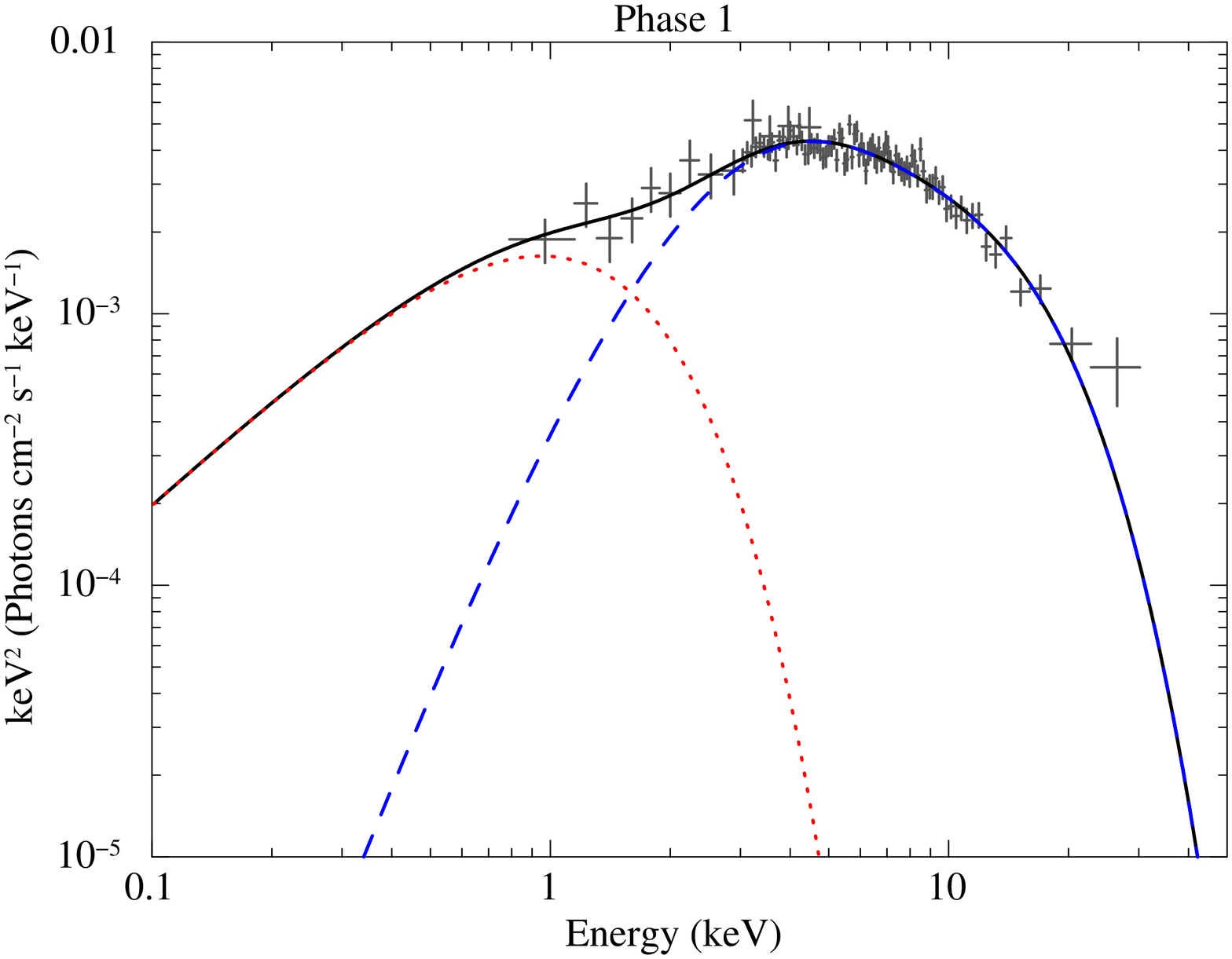}
\plotone{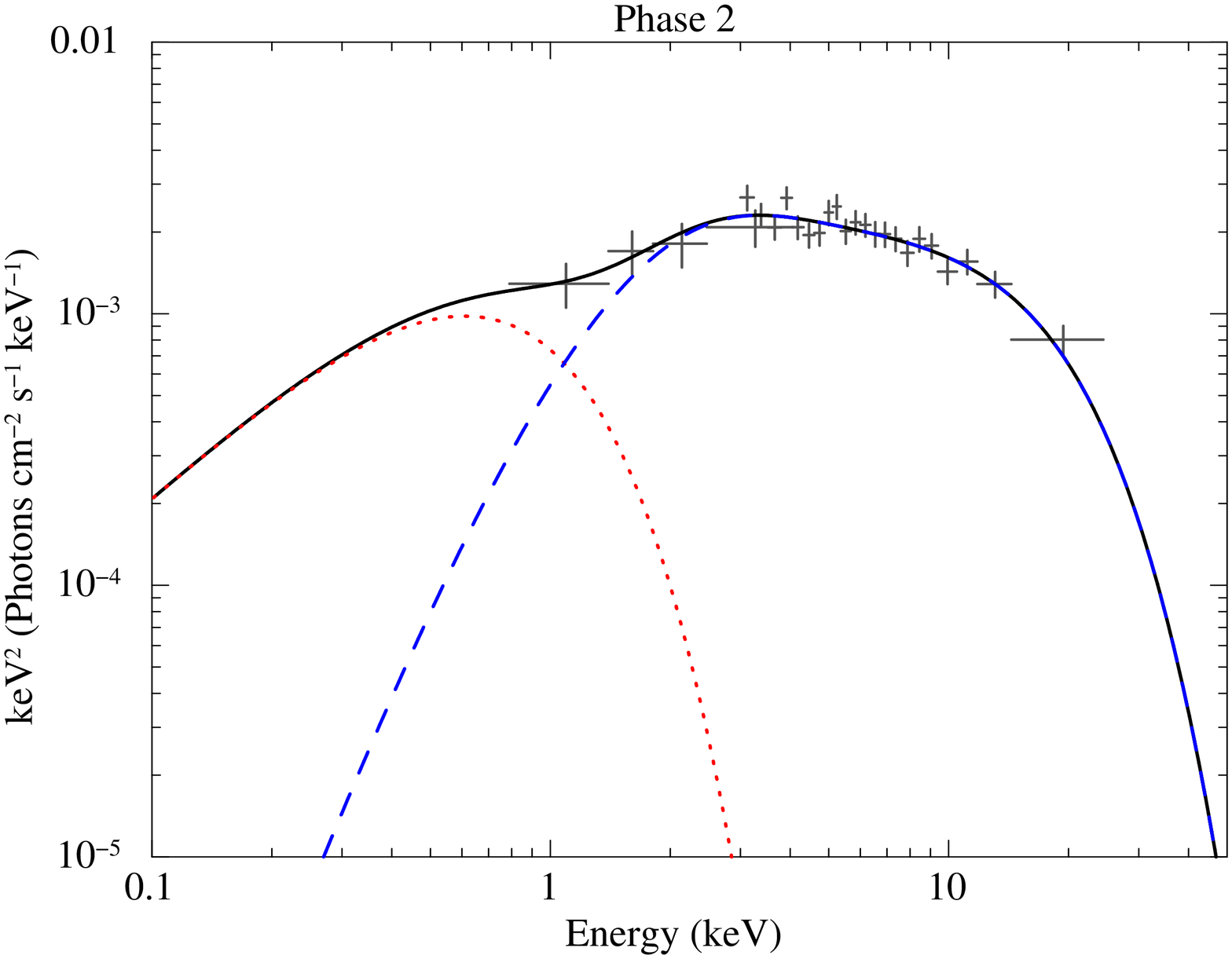}
\caption{
Absorption corrected best-fit {\tt diskbb+compTTm} models 
and data in 2012, Phase 1, and Phase 2 from top to bottom. 
The contributions of the {\tt diskbb} and {\tt compTTm} 
components are plotted in red dotted and blue dashed lines, 
respectively. Only the EPIC-MOS1 and {\it NuSTAR} data 
are shown in the top panel.
\label{fig_eeumodel_compttm}}
\end{figure}

\subsection{Comparison of the spectra in 2016 and 
2012 using a modified {\tt compTT} model}

\begin{deluxetable*}{lcccc}
\tablecaption{Best-fit parameters of {\tt diskbb+compTTm}  
models for the 2012 data and the 2016 data in Phase 1 and 
Phase 2
\label{tab_specfit_compttm}}
\tablewidth{0pt}
\tablehead{
\colhead{Parameter} & \colhead{Unit} & \colhead{2012} & \colhead{2016, Phase 1} & \colhead{2016, Phase 2} 
}
\startdata
$N_{\rm H}$ & $10^{22}$ cm$^{-2}$ & $0.37^{+0.05}_{-0.04}$ 
& $0.4^{+0.4}_{-0.4~{\rm (pegged)}}$ & $0.4$ (fixed) \\
$kT_{\rm in}$ & keV & $0.40 \pm 0.05$ 
& $0.40 \pm 0.07$ & $0.25^{+0.11}_{-0.7}$ \\
$N_{\rm bb}$ &  & $5^{+5}_{-2}$ 
& $9^{+18}_{-7} $ & $32^{+80}_{-32~{\rm (pegged)}}$ \\
$kT_{\rm 0}/kT_{\rm in}$ & & $2.0 \pm 0.1$ 
& $2.0$ (fixed) & $2.0$ (fixed) \\
$kT_{\rm e}$ & keV  & $3.7^{+0.6}_{-0.4}$ 
& $3.3^{+1.3}_{-0.5}$ & $3.9^{+2.0}_{-0.7}$  \\
$\tau$ & keV  & $5.1^{+0.7}_{-0.8}$ 
& $4.2^{+1.0}_{-1.4}$ & $4.2^{+1.0}_{-1.3}$ \\
$N_{\rm compTT}$ & & $(1.6 \pm 0.3) \times 10^{-4}$ 
& $(8 \pm 3) \times 10^{-4}$ & $(5 \pm 3) \times 10^{-4}$\\
$\chi^2/$d.o.f. & & $699/712 = 0.98$ 
& $89/91 = 0.97$ & $17/21 = 0.79$ \\
Flux & $10^{-11}$ erg cm$^{-2}$ s$^{-1}$ & $0.57$ 
& $1.7$ & $1.1$ \\
Luminosity & $10^{40}$ erg s$^{-2}$& $1.0$ 
& $3.1$ & $2.0$ 
\enddata
\tablecomments{The flux and luminositys listed above are unabsorbed 
values estimated in the 0.3--30 keV band. In all models, we included 
an additional {\tt TBabs} component with $N_{\rm H} = 3.0 \times 10^{21}$ cm$^{-2}$ 
as the Galactic absorption, which is omitted in the table.}
\end{deluxetable*}

In the previous sections, we assumed that 
the inner disk temperature of the {\tt diskbb} 
component is the same as the temperature 
of seed photons for the {\tt compTT}
component.
However, the seed temperature could be higher 
than the inner temperature determined from the 
direct MCD component, if the inner disk region 
is covered by optically-thick Comptonizing cloud 
and the pure MCD emission is only visible outside it. 
To investigate if the two temperatures
can differ, we attempt to model the spectra in 
2012 and 2016 varying $T_0$ and $T_{\rm in}$ 
independently. For this 
purpose, we modified the code of {\tt compTT} 
so that  $T_0$ is internally calculated from 
the two input parameters, $T_{\rm in}$ and 
$T_{\rm 0}/T_{\rm in}$. This model, that hereafter 
we call {\tt compTTm}, enable to vary $T_{\rm 0}$ 
and $T_{\rm in}$ keeping the condition 
$T_{\rm in} < T_{\rm 0}$.

We first fit the 2012 spectrum with the 
{\tt diskbb+compTTm} model, where we link
$T_{\rm in}$ of the two components and set 
the lower limit of $T_{\rm 0}/T_{\rm in}$ to 
be 1. The model gives an acceptable fit 
with $T_{\rm 0}/T_{\rm in}=2.0 \pm 0.1$.
The resultant parameters are listed 
in Table~\ref{tab_specfit_compttm}.
We note that the $\chi^2$ value is 
slightly improved from the {\tt diskbb+compTT} 
model (where $T_{\rm 0}/T_{\rm in} = 1$).
The best-fit model and data are shown in 
Figure~\ref{fig_eeumodel_compttm}.

Next, we apply the same model to the spectra 
in 2016. For phase 1, we get an unreasonably 
large electron temperature and small optical 
depth when we vary $T_{\rm 0}/T_{\rm in}$. 
We thus fix it at the best-fit value for the 
2012 data (2.0) to fit the spectra in two phases. 
As in Section~\ref{subsec_phase2_2012_fit}, we 
fixed $N_{\rm H}$ for Phase 2 at the value 
estimated from the Phase 1 data, to constrain 
the other parameters. The estimated parameters 
are given in Table~\ref{tab_specfit_compttm} 
and the best-fit models and the spectra for the 
two phases are presented in 
Figure~\ref{fig_eeumodel_compttm}.
The model with $T_{\rm 0}/T_{\rm in} = 2$ well fit 
both spectra. 
Remarkably, they are described with similar 
parameters to those estimated from the 2012 
data, except for the normalization of {\tt compTTm}, 
which is largest in Phase 1.

\section{Discussion} \label{sec_discussion}

Our X-ray observing campaign in 2016 gave an opportunity 
to study spectral evolution of IC 342 X-1. 
The coordinated {\it NuSTAR} and {\it Swift} 
observations performed in October 16--17 successfully 
captured the very moment that the ULX was changing its 
spectral shape, and provided the broad-band X-ray data 
during the spectral variation. In Phase 1, 
when the source marked the highest 3--20 keV flux in our 
campaign, the spectral shape appeared to be a more 
convex shaped than those observed with {\it XMM-Newton} 
and {\it NuSTAR} in 2012 \citep{ran15} and with {\it Swift} 
in 2016 October 9--11. 
The Phase-1 spectrum looks like the ``soft state'' 
observed with {\it ASCA} \citep{kub01, miz01} and with 
{\it Chandra} \citep{mar14}. Our {\it NuSTAR} observation 
have provided, for the first time, the hard X-ray spectrum 
of IC 342 X-1 above 10 keV in this state. We found that 
the observed spectrum deviates significantly at high 
energies from the profile of the single MCD model, 
demonstrating the importance of hard X-ray data. 
Our observations indicate that the transition to the 
broad convex spectrum takes 1 week or less and the 
transition back to the flatter spectrum took several hours 
or more. These timescales are a similar order of magnitude 
to the state transitions in Galactic BHXBs. 

As the X-ray luminosity decreased, the spectral 
profile became flatter and the location of the 
spectral turnover, which is most likely produced by 
thermal Comptonization, shifted to higher energies. 
Although this is clearly seen in 
Fig.~\ref{fig_compare_spec}, it is difficult to 
quantify the difference with the best-fit model.
Referring the results from the phenomenological 
cutoff power-law model that roughly approximates 
the spectrum in each epoch, we find that the photon 
index and cutoff energy in Phase 2 are significantly 
larger than those estimated in Phase 1 but 
are consistent with those in the 2012 epoch. 
This suggests that the 
properties of the accretion flow (and outflows) 
approached what was observed in the 2012 observation. 
The cutoff energy estimated for Phase 1 is about 
a factor of two smaller than the values in 2012 and 
in Phase 2, when the source had a factor of 2--4 
lower luminosity. Similar anti-correlations between 
the energy of the spectral turnover and the X-ray 
luminosity were found in previous observations of 
the same source \citep{yos13} and other sources 
such as Holmberg IX X-1 \citep{wal14},  
NGC 1313 X-2 \citep{pin12}, and NGC 5204 X-1 
\citep{pin14}. 

All the spectra in 2016 and 2012 are best 
reproduced with an MCD and a Comptonization 
model. We have investigated the difference 
between the inner disk temperature of the 
MCD component ($T_{\rm in}$) and the 
temperature of seed photons for the 
Comptonization component ($T_{\rm 0}$), 
which was often ignored in previous studies. 
We have found that the 2012 spectrum is best 
described with $T_{\rm 0}/T_{\rm in} = 2.0$ 
and that the two spectra in 2016 are 
also well characterized with the 
same temperature ratio and similar values 
of the other parameters to those 
obtained from the 2012 spectrum.
This may mean that what we have called as 
state transition between the soft state 
and the power-law state in IC 342 X-1 is, 
in reality, not a discrete change in the 
structure of the accretion flow but small 
variation in its properties caused by a 
smooth change in the mass accretion rate.
Common characteristics of the Comptonization 
component in ULXs are evidently present 
in our results: a low electron 
temperature ($T_{\rm e} \lesssim 10$ keV) 
and a large optical depth ($\tau \gtrsim 3$).

The electron temperature estimated in IC 342 X-1 
spectra, as well as those in many other ULXs 
\citep[below $\approx 10$ keV; e.g.,][]{gla09, bac13, wal13}, 
are much lower than the typical value of the low/hard state 
($\sim$100 keV) seen in transient BHXBs in our galaxy 
when their luminosities are below 1--10 \% of Eddington 
luminosity \citep[see, e.g.,][for details]{mcc06, don07}. 
An intermediate mass black hole above 
$\sim$1000 $M_\sun$ accreting at 
a sub-Eddington rate is thus unlikely as the central 
compact object in IC 342 X-1, with a typical luminosity 
of $\sim 10^{40}$ erg s$^{-1}$. 
Lower electron temperatures than in the low/hard 
state were obtained in Galactic BHXBs at higher 
luminosities. Comptonization components with $T_{\rm e} 
\approx$20 keV have been detected in several sources 
\citep[e.g.,][]{kub04, kub04b, tam12, hor14}, 
when they are in the "very high state", 
observed above a few ten percent of Eddington 
luminosity. Even smaller temperatures 
for the Comptonization component, similar to 
those of ULXs, are obtained in the anomalous 
"ultrasoft state" of GRO J1655$-$40 
\citep{utt15, shi16, nei16} and 
GRS 1915$+$105 \citep[e.g.,][]{ued10, nei11} 
at around Eddington luminosity. 

One possible origin of the cool, optically thick 
Comptonization component  
is dense gas bound to the inner disk. \citet{gla09} argued 
that an extreme corona, with a low electron temperature 
and high optical depth, covers the inner disk region. 
Several recent works have suggested an alternative 
possibility that the observed spectral 
components are produced by emission from an 
inhomogeneous inner disk 
\citep{mil13, mil14, wal13, ran15}, which could form 
due to an instability caused by strong radiation 
pressure \citep{dex12}.  
A distorted thermal spectrum originating in such disk 
could describe the spectral profile in ULXs as well 
as that in the very high state of Galactic BHXBs. 
A power-law tail in addition to the thermal 
Comptonization component is detected previously with 
{\it NuSTAR} in IC 342 X-1 \citep{ran15} and a few ULXs 
\citep{wal13, wal14}, 
which could be associated with the steep power-law component 
dominating the hard X-ray flux in the very high state. 
We note, however, that the absorption lines seen in 
GRS 1915$+$105 and GRO J1655$-$40 in the ultrasoft state 
have one of more order(s) of magnitude 
smaller values of the blueshifts (typically $\sim 0.001 c$; 
e.g., \citealt{ued09}, up to $\sim 0.01c$ 
\citealt{mil15}) than that recently discovered in a ULX 
\citep[$\approx 0.2 c$,][see below]{pin15, wal16}. 
This may suggest that the accretion flows of ULXs 
have different physical properties 
from those of Galactic BHXBs below or at around 
Eddington luminosity.

%% comparison with numerical simulations
An alternative interpretation is that the Comptonization 
component is produced by a powerful radiation-driven 
outflow predicted to launch from super-Eddington 
accretion flows \citep{ohs05, ohs11}. Recent observations 
have provided evidence that such a wind does exist 
in ULXs. Absorption features have now been discovered 
at $\approx 1$ keV \citep{pin15} and $\approx 9$ keV 
\citep{wal16} in the ULX NGC 1313 X-1, which confirmed 
the previous suggestion that the complex spectral 
structures often seen in ULXs may be linked to disk 
winds at a velocity of $\sim 0.1 c$ \citep{mid14}. 
Also, recent studies of ultraluminous soft X-ray 
sources (ULSs), dominated by thermal component at 
a temperature of $\sim 0.1$ keV, found an association 
between ULXs and ULSs and successfully described the 
X-ray spectra of both classes with a Compton-thick 
winds with different optical depths \citep{sor16, urq16}.

The Compton-thick outflow model naturally explains 
the behavior of the spectral turnover that we observed. 
Calculating spectra based on the results of radiation 
hydrodynamic simulations, \citet{kaw12} showed 
that the turnover is seen at lower energies at higher 
luminosities. As shown in Figure~2 of \citet{kaw12}, 
the energy of the turnover increases by a factor of 2 
from $\sim 5$ keV to $\sim 10$ keV and the 
overall spectral profile become somewhat flatter, 
when the mass accretion rate decreases from 
$\sim 5 \times 10^2 L_{\rm Edd}/c^2$ to $\sim 2 \times 10^2 L_{\rm Edd}/c^2$.
This is consistent with the 
observed spectral variation. The electron temperature 
could decrease due to stronger Compton cooling 
at higher mass accretion rates. Although in our best-fit 
{\tt compTTm+diskbb} results, its error ranges for 
the individual spectra overlap one another, future 
observations may give better constraints by providing 
data with better statistics in the hard X-ray band 
and enable us to discuss the variation of $T_{\rm e}$.

If the Comptonization component in IC 342 X-1 is 
produced by a Compton thick outflow, 
the soft component, which we modeled with 
{\tt diskbb}, may originate in the accretion disk 
outside it or in the photosphere in the outer 
regions of the outflow itself \citep[e.g.,][]{pou07}. 
If the former is the case, the inner radius derived from 
the direct MCD component could be regarded as the 
boundary between the Compton thick and Compton thin 
regions of the outflow. The normalizations of the {\tt diskbb} 
components obtained from the {\tt diskbb+compTTm} model 
give $R_{\rm in} = 9^{+4}_{-2} \times 10^2~(\cos i/\cos 0^\circ)^{-1/2}$ km 
(for the 2012 observation, where $i$ is the inclination angle), 
$R_{\rm in} = 1.2^{+0.9}_{-0.6} \times 10^3~(\cos i/\cos 0^\circ)^{-1/2}$ km 
(Phase 1), and $R_{\rm in} < 4.2 \times 10^3~(\cos i/\cos 0^\circ)^{-1/2}$ km 
(Phase 2), where the correction factor for the spectral hardening and 
the boundary condition is assumed as unity \citep{kub98}. 
According to latest numerical simulation of 
super-Eddington accretion flows, the outflows are 
Compton thick only within $\approx$ 20 $R_{\rm g}$\footnote{$R_{\rm g}$ 
is the gravitational radius ($GM_{\rm BH}/c^2$, where $G$, $c$, $M_{\rm BH}$ 
are the gravitational constant, the light speed, and the black 
hole mass).} 
(\citealt{kaw12}; Kitaki et al.\ in preparation). If the 
black hole mass of IC 342 X-1 is $\sim 30 M_\sun$, the radii 
that we estimated are consistent with the prediction 
of the simulation.

\section{Conclusions}
We performed an X-ray observing campaign of IC 342 X-1 
with {\it NuSTAR} and {\it Swift} in 2016 October, which 
allowed us to study spectral evolution. We have found 
that the cool $T_{\rm e} \approx 4$ keV, optically-thick 
($\tau \approx 5$) Comtonization exists not 
only in the low-luminosity state but also in the 
``soft state'' seen at high luminosities, where the 
spectrum exhibits a broad convex shape. 
We have found that the spectral components and their 
variation can be interpreted in the context of the 
super-Eddington accretion flow with Compton-thick 
outflows around a stellar-mass black hole, and that 
the spectral variability in IC 342 X-1 
can be explained by a continuous change in mass 
accretion rate. 

\acknowledgments

We are grateful to the {\it NuSTAR} and {\it Swift} operation teams 
for carrying out the ToO observations. 
We thank Kirill Atapin for helpful comments, and 
Ken Ohsuga and Tomohisa Kawashima for useful 
discussion of latest results from numerical simulations. 
MS acknowledges support by the Special Postdoctoral Researchers 
Program at RIKEN. This work is partly supported by a Grant-in-Aid 
for Young Scientists (B) 16K17672 (MS) and for Scientific Research 
26400228 (YU). SF acknowledges support by the the Russian RFBR 
grants 16-02-00567, 15-42-02573, and the Russian Science Foundation 
grant 14-50-00043. This research has made use of data supplied by 
the UK Swift Science Data Centre at the University of Leicester, 
and data obtained from the High Energy Astrophysics Science Archive 
Research Center (HEASARC), provided by NASA's Goddard Space Flight Center.

\end{document}